%% file: HST-1.tex
\documentclass[cits]{PoS}
\usepackage{amsmath,amssymb,amsbsy,amsfonts}
\usepackage{url}
\input{aas-journals}

\title{Understanding the light curves of the HST-1 knot in M87}

\ShortTitle{Understanding the light curves of HST-1}

\author{{Y. U. Coronado} and S. Mendoza \\
        Instituto de Astronomia, Universidad Nacional Autonoma de Mexico.\\
        E-mail: \email{coronado@astro.unam.mx},
                \email{sergio@astro.unam.mx}}

\abstract{The motion of knots in astrophysical jets is commonly
interpreted as shock waves moving along it. 
Observations of the HST-1 knot during extended periods of time
have produced complicated light curves on many wavelengths which are
difficult to account using standard hydrodynamical models.
Here we reproduce these light curves using the 
semi-analytical approach developed by Mendoza et al. (2009),
developed to reproduce light curves of working surfaces moving along
relativistic jets.  These working surfaces are generated by periodic
oscillations of the injected flow velocity and discharge
at the base of the jet.  In particular,
we use this approach to reproduce the exotic observed features of the
light curves of the HST-1 knot in M87.  We show that the complicated fits
to the light curves are reproduced with high accuracy in all wavelengths.
As a bonus, we show that this model is also able to reproduce the light
curve of the micro-qsr A06200-00 with high accuracy.}

\FullConference{25th Texas Symposium on Relativistic Astrophysics - TEXAS 2010\\
		December 06-10, 2010\\
		Heidelberg, Germany}

\begin{document}
      
\section[Introduction]{Introduction}

 Chandra Observations of the HST-1 knot \cite{harris06}
in M87 showed that a peak in the X-ray light curve developed about 2005.
This light curve has since shown successive peaks over short periods of
time.  Observations in UV \cite{madrid09} and radio \cite{chang10}
have also shown the same trend.  Since quasi-periodic signatures in the
brightening and dimming of the HST-1 knot X-ray observations were found
\cite{harris09}, this shows a manifestation of a previous modulation
in the jet power, most probably a local oscillation of the process that
converts the bulk kinetic jet power to internal energy of the emitting
plasma.

 The key to understand the particular features in the HST-1 knot of M87
is the physical basis behind its light curve.  To give an idea of the
complexities observed, there has been a discussion about the X-ray
observations as to whether the main contribution in this wavelength is an
effect of the hot accretion disc with its corona \cite{cheung07} or a
particular feature of some kind of recollimation shock \cite{Stawarz06}.  We
mention here that radio observations show that the knot is well isolated
from the nucleus since the activity is displaced away from the central
engine by \( \gtrsim 120\, \text{pc} \).  Also, these radio observations
show superluminal motion of the knot, which means that its bulk velocity is
highly relativistic.

  In this article we take multi-wavelength observations from
different sources. For instance, the X-rays observations are part of
a multi-frequency program coordinating the Chandra and HST monitoring
by \cite{harris09}. The ultraviolet data are part of the observations
carried out during the years 1999 to 2006 \cite{madrid09}. Finally the
radio data corresponds to observations with the VLBI at \( 2 \, \text{cm}
\) \cite{chang10}.  All these observations reveal a clear multi frequency
light curve of the HST-1 knot which serves as a laboratory to test the
ideas developed by \cite{mendoza09} for the formation and propagation
of the working surface of a relativistic jet, with periodic variations
of the injected velocity profiles and mass rate outflows.

\section[model]{Model}

 The formation of internal shock waves on a relativistic jet are
commonly explained by different mechanisms, such as the interaction of
the jet with inhomogeneities of the surrounding medium, the bending
of jets and time fluctuations in the parameters of the ejection
\cite{rees94,mendoza-phd,jamil08,mendoza09}.  Here we are concerned
with the latter. When the speed of the emitted mass particles varies
with time, a faster but later fluid parcel eventually hits an earlier
but slower ejection producing an initial discontinuity which gives rise
to a working surface, i.e.  a hydrodynamical object formed by two shock
waves separated by a contact surface.  In the frame of reference of the
central engine, where the jet is being ejected, the working surface
travels along the jet with an average velocity \( v_\text{ws} \).
(see e.g. \cite{rees94}).  In what follows we use the relativistic
semi-analytical model of \cite{mendoza09} to describe a working surface
and its kinetic luminosity power travelling inside an astrophysical jet.
To do this, we consider a source ejecting material in a preferred
direction \( x \) with a velocity \( v(\tau)  \) and a mass ejection
rate \( \dot{m}(\tau) \), both dependent on time \( \tau \) as measured
from the jet's source \cite{mendoza09}.  The energy loss inside the working
surface is calculated as the difference between the initial energy of the
material, when it was injected at the base of the jet, and the energy of
the flow inside the working surface. Assuming an efficient mechanism which
converts all this kinetic energy into radiation power, then the total
luminosity is given by \( L := \mathrm{d} E_\text{r} / \mathrm{d} t \),
where \( E_\text{r} = E_0 - E_\text{ws} \) is the radiated energy within
the working surface \cite{mendoza09}.

  In what follows we use the approach followed by \cite{mendoza09}
in order to show that their model can describe the important features
observed on the evolution of \( HST-1 \).  To do so, the injected
velocity is assumed to have a periodic variation given by \( v(\tau)
= v_0 + c \eta^2 \sin \omega \tau \), with a constant mass discharge
\( \dot{m} \) and a velocity of light \( c \).  We mention here that in
order to perform the computations it is best to numerically assume a
dimensionless system of units for which the oscillating frequency \(
\omega = 1 \) and \( \dot{m} = 1 \).  We have used this approach, which
is also followed by \cite{mendoza09}, but we will present our results
on the physical system of units for which \( \omega \) and \( \dot{m} \)
are dimensional constants.

\section[fit]{Numerical fits}

  The data sample of HST-1 covers a period of time between the years 2000
to 2009.  Since we are going to use observations on different wavelengths,
then it is best to normalise all observations to the intrinsic luminosity
of HST-1.  To do so, for X-rays we use the procedure developed in
\cite{harris06}, which gives a power law index for the flux of \( 1.5 \)
at a wavelength of \( 2 \textrm{cm} \) \cite{chang10}.  For the UV data
we use the flux density using the reference wavelength of the camera
ACS/F220W at \( 2255.5 \textrm{AA} \) \cite{madrid09}.

  Since the correction in flux is given, we can now use the standard
relation between luminosity and flux without worrying about extinction.
To do so, we assume an isotropic emission of the source at a distance of
16 Mpc \cite{Jordan05}, which gives a lower limit in the luminosity emitted
by the HST-1 knot.  

  On the other hand, the mean velocity \( v_0 \)  of the
jet is taken as \( v_0 = 0.98 c \) which is in agreement with the observed
value of a Lorentz Factor \( \gamma = 6 \) \cite{biretta99}.
The value of the parameter \(\eta^2\) and frequency are selected by linear
fits in light curve of the observed data.

 The X-ray light curve does not work with a simple variation of the
velocity and so, we additionally adopt a periodic variation on the injected
mass given by $ \dot{m}  = \dot{m}_i + \psi \sin \Omega$ just for the first
flare.  After the peak, the standard assumption made above fits quite well
the observations.  

 In all the observed light curves, there is at least one subsequent
increase in luminosity after the maximum peak.  These local peaks can be
easily modelled by assuming a rapid variation on the value of 
the discharge $\dot{m}$ injected in the jet according to the 
high variability of the core of M87 \cite{harris09}.

\begin{figure}
\begin{minipage}[b]{0.5\linewidth}
\centering
 \includegraphics[scale=0.60]{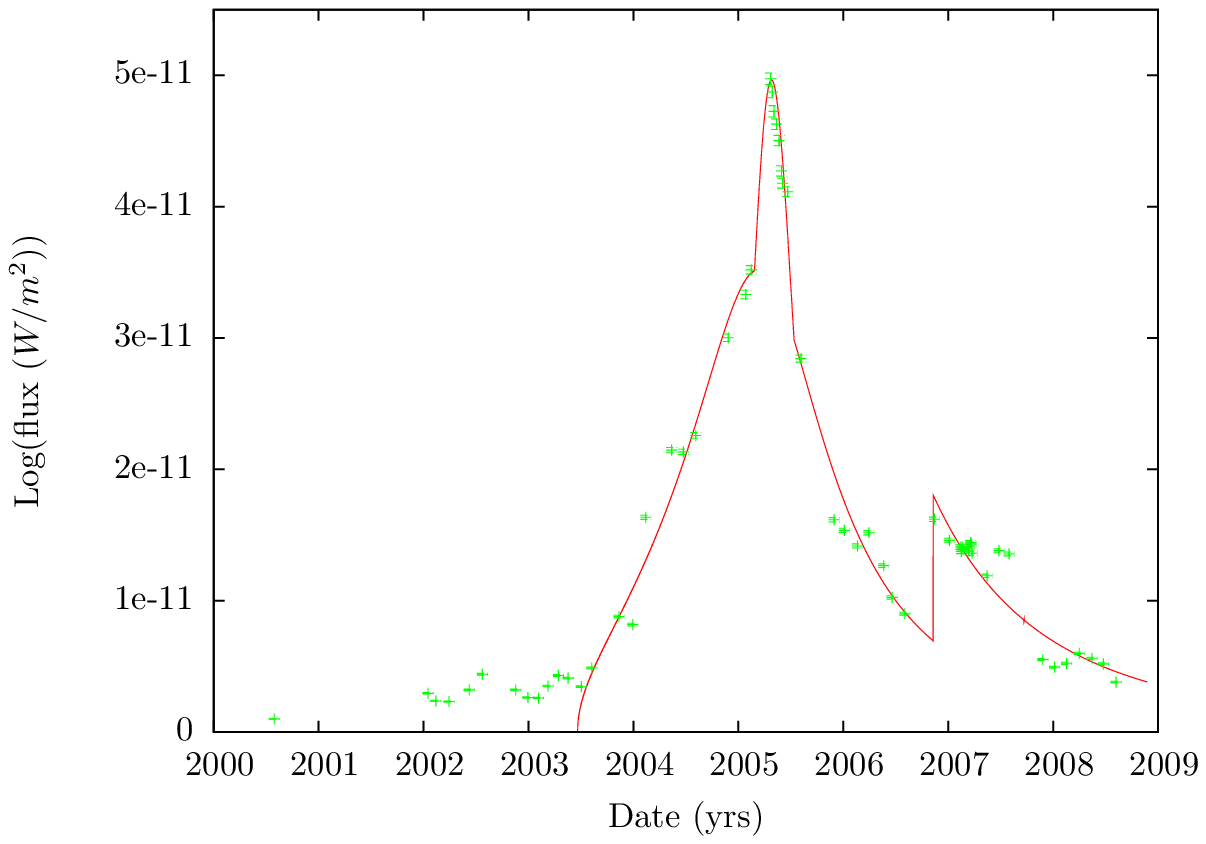}
\caption{The figure shows as points the observed light curve in X-Ray
of HST-1 from \cite{harris09}. The continuous curve on the plot is the
best fit using the semi-analytical model described in the text with
a linear fit to the data yielding \( \dot{m} = 5.59 \times 10^{-7} M_{\odot}/yr \) and
\( \omega = 0.1207 \).  The second peak after the maximum is modelled as an 
increase in the discharge of \( \dot{m} = 1.10 \times 10^{-7} M_{\odot}/yr\) at time
\( 2006.86 \) with a duration of \( 1.64 \) years.}
\label{fig01}
\end{minipage}
\hspace{0.5cm}
\begin{minipage}[b]{0.5\linewidth}
\centering
 \includegraphics[scale=0.60]{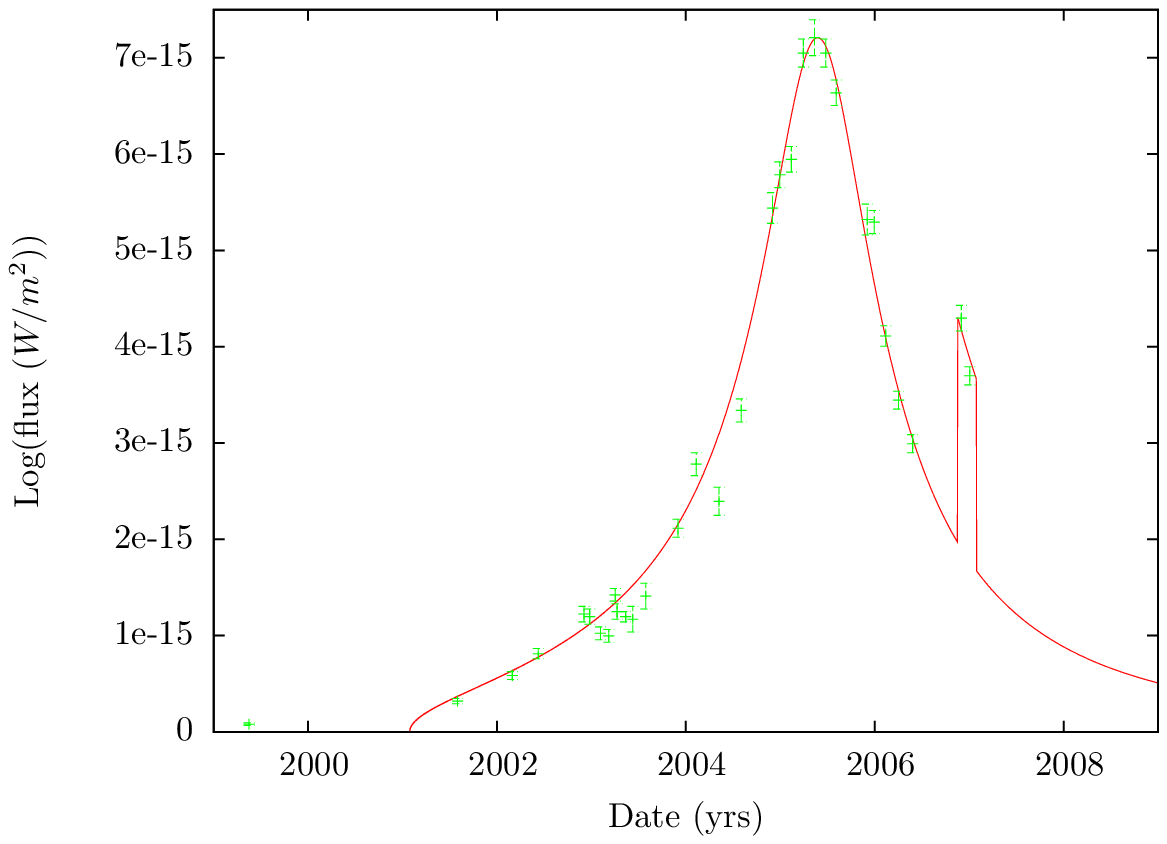}
\vskip 5.0mm  
\caption{The figure shows the observed points of the HST-1 UV light curve
from \cite{madrid09}. The continuous curve on the graph is the best fit to
the observations using the semi-analytical model described in the text
with a linear fit to the data yielding  \( \dot{m} = 6.52 \times 10^{-6} M_{\odot}/yr\)
and \( \omega = 0.02528 \).  The second peak after the maximum is modelled as an
increase in the discharge of \( \dot{m} = 1.78 \times 10^{-6} M_{\odot}/yr\) at time
\( 2006.88 \) with a duration of \(0.2 \) years. }
\label{fig02}
\end{minipage}
\end{figure}

\begin{figure}
\begin{minipage}[b]{0.5\linewidth}
\centering
 \includegraphics[scale=0.65]{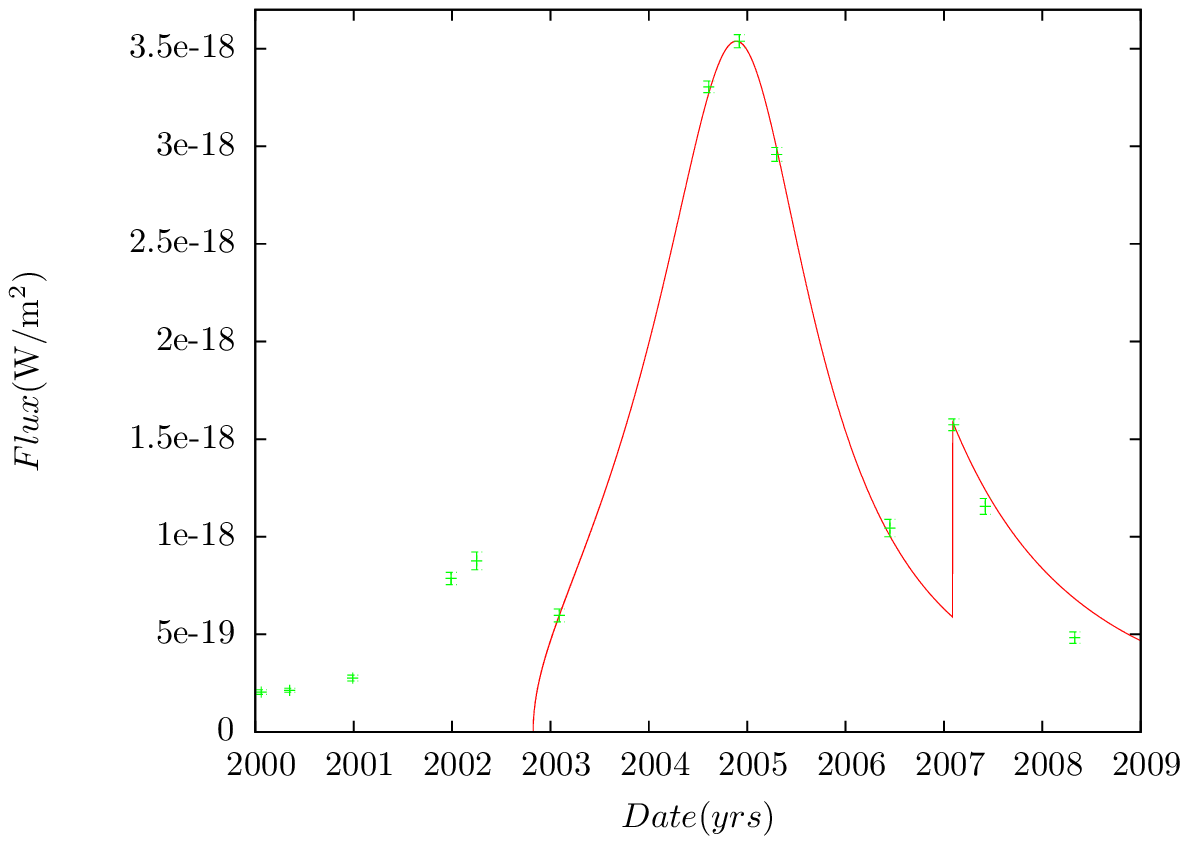}
\caption{The figure shows the observed points of the HST-1 radio
light curve from the observations of \cite{chang10}.  The continuous curve
on the figure is the best fit to the observations using the semi-analytical
model described in the text with a linear fit to the data yielding  \(
\dot{m} = 2.11 \times 10^{-11} M_{\odot}/yr \) and \( \omega = 1.0 \).  The second peak after 
the maximum is modelled as an increase in the discharge of \( \dot{m} = 3.48
\times 10^{-12} M_{\odot}/yr\) at time \( 2007.1 \) with a duration of \( 2.0\) years. }
\label{fig03}
\end{minipage}
\hspace{0.5cm}
\begin{minipage}[b]{0.5\linewidth}
\centering
 \includegraphics[scale=0.65]{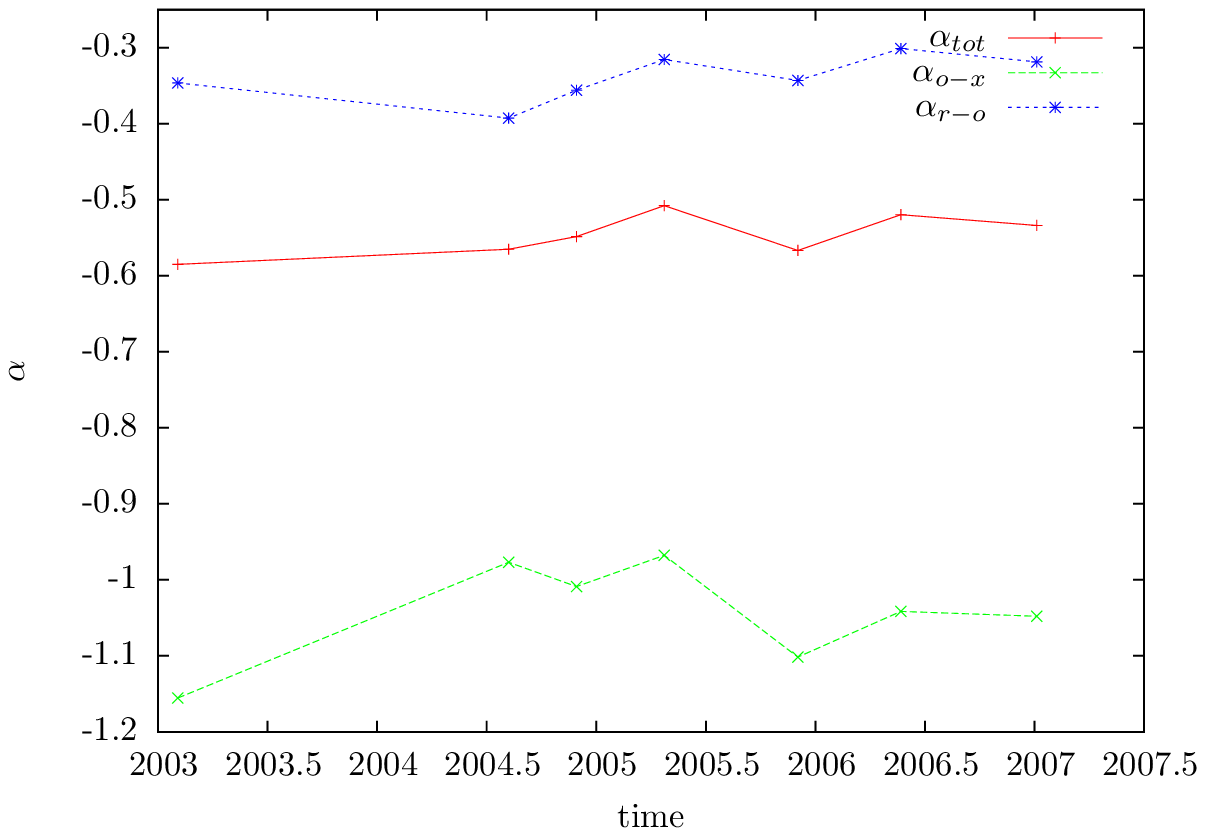}
\vskip 14.0mm 
\caption{The figure show the time \( t \) evolution of the spectral index
\( \alpha \).  The upper and lower curves are the spectral values from
radio to UV (\(2 \text{cm}\)to \(225.5 \text{nm}\)) and from UV to X.rays
(\(225.5 \text{nm}\) to \( 2 \text{KeV}\)) respectively.  }
\label{fig04}
\end{minipage}
\end{figure}

 Figures~\ref{fig01}-\ref{fig03} show the obtained fits, which are in good
agreement with the observations.  However, all of these take into account
the bolometric luminosity of the flow.  We have calculated in
Figure~\ref{fig04} the variation of the spectral index as a function of time
to fit our model, with data obtained by \cite{harris06}.
From the figure it follows that the X-ray luminosity is dominant at all
times.


\section{Discussion}

  We have described the light curve of the knot in HST-1 on different
wavelengths using the relativistic model of the evolution of working
surfaces moving along a jet by \cite{mendoza09}.  The physical mechanism
responsible of the emission is still not known since, as described by
\cite{harris06}, a simple synchrotron emission can't be accounted due to a
break in the main burst.

\begin{figure}
\begin{center}
 \includegraphics[scale=0.65]{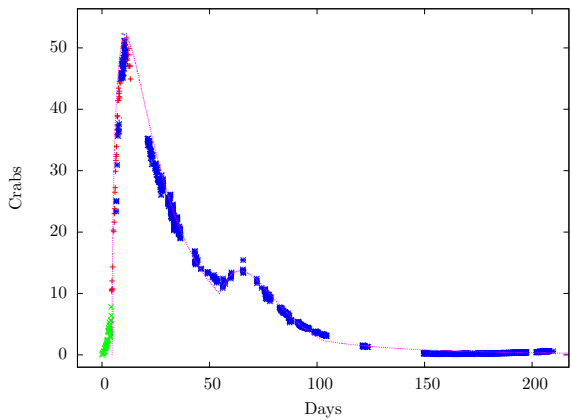}
\end{center}  
\caption{The figure show a burst on the X-ray light curve of the
microquasar A06200-00.  The continuous curve on the plot is the
best fit using the semi-analytical model described in the text with
a linear fit to the data yielding \( \dot{m} = 3.27 \times 10^{-17} M_{\odot}/yr\)
and \( \omega =1.0 \).  The second peak after the
maximum is modelled as an increase in the discharge of \( \dot{m} = 7.2
\times 10^{-14} M_{\odot}/yr\)  at time \( 55\) with a duration of \( 45 \) days.}
\label{fig05}
\end{figure}

 To see the power of the semi-analytic model of \cite{mendoza09}, we 
also adopt the same procedure in Figure~\ref{fig05} but for the \(\mu
\)-quasar A0620-00 which has been observed in detail for quite a nice
period of time.

  Since the same model has been used to fit light curves of long gamma-ray
bursts by \cite{mendoza09}, all this means that light curves that produce
bursts can be easily modelled by using this approach and that certainly the
burst are simply two shock waves forming due to the variations of the
injected flow.

\acknowledgments
This work was supported by a DGAPA-UNAM grant (PAPIIT IN116210-3).
The authors acknowledge support from CONACyT (210965,26344) and thank
J. McClintock for providing the observations of the \(\mu\)-qsr A0620-00.






\end{document}

%% file: aas-journals.tex
\newcommand{\aj}{Astronomical Journal}

\newcommand{\apj}{ApJ}
\newcommand{\apjl}{ApJL}

\newcommand{\mnras}{MNRAS}